\documentstyle[preprint,aps,epsfig]{revtex}

\newcommand \beq{\begin{eqnarray}}
\newcommand \eeq{\end{eqnarray}}

\begin{document}
\normalsize
\title{Probing the Phase Boundary between Hadronic Matter and the
Quark-Gluon-Plasma in Relativistic Heavy Ion Collisions}
\author{
P.~Braun-Munzinger$^1$ and J.~Stachel$^2$
}
\address{$^1$ Gesellschaft f\"ur Schwerionenforschung, Darmstadt, Germany\\
$^2$ Physikalisches Institut der Universit\"at Heidelberg, Germany}
\date{\today}
\maketitle

\begin{abstract}
We discuss recent data on particle production with emphasis on the
degree of thermal and chemical equilibration achieved. The data are
interpreted in terms of a resonance gas model. The phase boundary
constructed between the resonance gas and the quark-gluon plasma is
shown to be very close to the deduced parameters characterizing the
hadronic fireball at freeze-out.
\end{abstract}

\vspace{1cm}

\narrowtext

Collisions between ultra-relativistic heavy nuclei create matter of high
initial energy and particle density. The experimental program currently
underway at the fixed target machines probes at the same time the region
of very high initial baryon density. Under such conditions it is
expected that a phase of deconfined quarks and gluons, where also chiral
symmetry is restored, is more stable than hadronic matter. An
experimental heavy ion program has been underway for nearly ten years at
the 
Brookhaven AGS and the CERN SPS. In a number of experiments many
hadronic observables have been measured, often over a large fraction of
the full solid angle \cite{qm95}. With these experimental data  the final
freeze-out stage has been 
well characterized in terms of the hadronic observables at least for the
relatively light Silicon and Sulfur beams and similar data for the
heaviest beams, Gold and Lead, are rapidly becoming available. We want
to address in this article the question to what extent the final
hadronic stage can be characterized in terms of the concepts of thermal
and hadrochemical equilibrium, {\it i.e.} to what extent we can talk about a
thermodynamic phase. This will lead to the next question, namely which
regions of the phase diagram  are probed by present experiments. We
would like to point out that such studies of freeze-out configurations
do not shed direct light on the high density/high temperature phase of the
collision. On the other hand, if equilibrium is not reached at
freeze-out, it is difficult to see how one can describe the fireball
formed in the collision in terms of thermodynamical quantities at
earlier stages. 

As a starting point for the present discussion we use results from
recent numerical solutions of QCD on the lattice. For zero net baryon
density the critical temperature for the transition between the hadronic
and the quark-gluon phase has been obtained by several groups
\cite{lattice}: 
$T_c=145 \pm 10$ MeV. Lattice QCD calculations do not, up to now,  shed
light on 
the properties of the phase transition at finite baryon density. In
order to continue the lattice results into the region of finite baryon
density relevant for present fixed target experiments we construct \cite{thermal1,thermal2} the phase boundary by
equating chemical potential 
$\mu_B$ and pressure $P$ of a hadron resonance gas with the equivalent
quantities of 
a non-interacting quark-gluon plasma. The hadron resonance gas contains
all known baryons and mesons up to a mass of 2 and 1.5 GeV,
respectively. Interactions among the baryons are approximately taken into
account by an excluded volume correction \cite{thermal2}. The plasma
phase contains 
massless gluons, u and d quarks and strange quarks with $m_s$= 150
MeV. A bag constant of B = 262 MeV/fm$^3$ is used to insure that the  
transition for $\mu_B$ = 0 takes place at $T_c = 145 $ MeV, consistent
with the lattice QCD results. We note that, because of the relatively
low transition temperature, the hadron gas is never very dense near 
$T_c$ especially for $\mu_B < 0.8 $ GeV, {\it i.e.} the values relevant
for fireballs formed in ultra-relativistic nucleus-nucleus collisions
(see below). The excluded volume corrections are consequently not very
large. Also the  
neglect of very heavy hadrons ( with masses $ > 2$ GeV) in the resonance
gas is of no consequence for 
the calculation of the phase boundary in the relevant region. 

In order
to illustrate our construction of the phase transition 
we present, in Figure 1, the dependence of pressure on
temperature for two values of the baryon chemical potential, $\mu_B =
0.17$ and 0.54 GeV, respectively. The pressure in the hadronic phase
(starting at $P = 0$ for $T = 0$) rises approximately $\propto T^6$
because of the resonances included in the hadron gas description, while
for the quark-gluon phase $P \propto T^4$ and, of course, $P = -B$ at $T=0$. 

Following the suggestion of
\cite{hung}, we plot, in Figure 2,  the
equation of state in a plot of pressure divided by energy density
vs. energy 
density, {\it i.e.} $P/\epsilon$ as function of $\epsilon$ for $\mu_B =
0.54$ GeV. Similar to the results obtained by \cite{hung} our equation
of state at finite baryon density also has a ``soft point'', {\it i.e.}
smallest pressure for a given energy density, at $\epsilon
\approx 1.2$ GeV/fm$^3$, although the detailed shape of the curve,
especially in the hadron gas region, is quite different from that of
\cite{hung}. We furthermore note that, at the softest point, the
pressure is about $P \approx 50 $ MeV/fm$^3$, not
negligible but small when compared to $\epsilon/3$. Also shown in Figure
2 is the    
dependence of the square of the speed of sound, $c_s^2$ on $\epsilon$. 
The speed of sound, $c_s = \sqrt{dP/d\epsilon}$ vanishes by
construction in the
mixed phase.  The relatively low 
pressure and small speed of sound at the ``softest'' point suggest the
possibility of a very long lived fireball.

The full  phase diagram resulting from the above described construction is 
shown in Figure 3 both in terms of the chemical potential and
in terms of the baryon density. The fact that the baryon densities on
the quark-gluon and on the hadron side differ as soon as they are finite
reflects the first order nature of the phase transition constructed
in this way. 
For $\mu_B < 0.8 $ GeV the transition temperature varies only slightly,
reaching a value of $T_c \approx 120$ MeV at $\mu_B = 0.8 $ GeV. For
larger $\mu_B$ values the position of the phase boundary becomes rather
uncertain because of the rather crude way in which we incorporate
baryon-baryon repulsion  through an excluded volume correction
\cite{thermal1,thermal2}. However, our analysis will focus on smaller
values of the baryon chemical potential, where baryon densities at
freeze-out are significantly below normal nuclear matter density. In
this region the interaction corrections are small and we consider the
phase diagram to be consistent within the framework chosen.

As a next step we demonstrate to what accuracy the hadron abundances  
from AGS and SPS experiments with Si and S beams can be described in
terms of the two thermodynamic variables $T$ and $\mu_B$. This is shown
in Figure 4 and Figure 5. The details of the analysis
are described in \cite{thermal1,thermal2}. The resulting values are
$\mu_B=0.54$ GeV, 
$120 \leq  T \leq 140$ MeV at AGS energy and $\mu_B=0.17 (0.18)$ GeV,
$T = 160 (170) $ MeV  The overall agreement between
the measured particle ratios and the thermal model predictions is rather
impressive, on the
20\% level or better  in most cases. Note that, because of the large baryon
chemical potential and relatively low temperature, the particle ratios at AGS
energy vary over six 
orders of magnitude, while the ratios measured at the SPS are much more
bunched together, consistent with a relatively small $\mu_B$ value and
higher temperature. 

The baryon density at freeze-out for both the AGS and
the SPS data is $\rho_B \approx $ 0.06/fm$^3$, while the corresponding pion
densities are $\rho_{\pi} \approx$ 0.08/fm$^3 $ (AGS, $T = 120$ MeV) and
$\rho_{\pi} 
\approx $0.30/fm$^3 $ (SPS, $T = 160 $ MeV). This may reflect the fact
that freeze-out at 
the AGS, where the pion/nucleon ratio is near unity, is determined by the
pion-nucleon and nucleon-nucleon dynamics 
and the associated large cross sections ($\sim 100$ mb) while the
small $\pi \pi $ 
cross section ($\sim 10$ mb) is the relevant quantity at SPS energies
where pions dominate nucleons by about 5 to 1 in the central region.
To within the accuracy of present data this analysis shows that the
hadronic freeze-out configuration is very close to thermal and chemical
equilibrium. In particular, the strangeness suppression which is well
known for data from nucleon-nucleon collisions is not observed in
nucleus-nucleus collisions. Rather, strangeness degrees of freedom are
close to the values expected for a hot hadronic system in chemical equilibrium.

Armed with these results one may predict the hadronic abundances which
should be produced in Au-Au collisions at future colliders such as RHIC
or LHC. For these predictions we merely set $\mu_B = 0$ and set $T =
 T_c(\mu_B = 0) = 145$ MeV. The results for spatial densities and
rapidity densities of hadrons are shown in Table 1. While the spatial
densities are 
a direct result of the thermal model calculations, predictions of the
rapidity densities imply fixing of the fireball volume. We have adjusted
this volume to V = 9524 fm$^3$, such that 
the total pion rapidity density is 2000, {\it i.e.} a value typical for
what is 
expected for experiments at RHIC. The results show that a thermally
equilibrated fireball will contain about 170 nucleons and anti-nucleons,
and the $K^+/\pi^+$ ratio will be about 19\%.

An intriguing question is raised by the rather surprising result
of chemical equilibration at freeze-out for strange particles. Assuming free
hadron-hadron cross sections for strangeness production one would expect
that very long lifetimes are needed to bring strange particles into
chemical equilibrium. One way to bring strangeness into
equilibrium is the possibility that a deconfined phase (quark-gluon
plasma) was formed during the course of the collision. If part of the large
strangeness content of a thermalized plasma is carried over into the
hadronic phase \cite{barz}, the large ratios observed for
strange/non-strange particle production yields are readily explained. 

An alternative but somewhat related option has been demonstrated by Ko,
Brown and collaborators
\cite{kogeb}. It is that the possible reduction of particle masses in
the hot and 
dense fireball leads to a strong enhancement in strangeness production
cross sections, and consequently to much more rapid equilibration. In
particular, the dependence on temperature and density of nonstrange
hadron masses is
parameterized in \cite{kogeb} as:
\beq
  \frac{m^*}{m}=[1-(\frac{T}{T_c})^2]^{(1/3)}[1-0.2\frac{\rho_B}{\rho_0}],
\eeq

while for kaons, 
\beq
 \frac{m_K^*}{m_K}=[1-(\frac{T}{T_c})^2]^{(1/3)}[1-0.2\frac{\rho_B}{\rho_0}]^{1/2}      
\eeq

If we, considering the above determined freeze-out parameters,  assume,
for the case of a fireball created at AGS energy, that 
$\langle \rho_B/\rho_0 \rangle \approx 2$ and $\langle
T/T_c \rangle \approx 0.8$, where the brackets imply average
over the lifetime of the fireball, we get $\langle m^*/m \rangle
=0.43$ for nonstrange hadrons and 0.55 for the
kaons. Especially for the $\rho \rho \rightarrow K \bar K$ cross section
such mass reductions lead to substantial increases and, consequently,
much more rapid thermalization. One should note, however, that the
question, how hadron and, in particular, kaon masses behave in dense
matter, is not settled. In 
particular, Lutz {\it et al.} \cite{lutz} come to rather different
conclusions for the in-medium mass change, with the K$^+$ mass actually
increasing with increasing baryon density.

Nevertheless, if chemical equilibration of the
strangeness degrees of freedom at freeze-out
is 
confirmed for the heavy systems such as Au+Au at the AGS and Pb+Pb at
the SPS, this could be considered evidence, albeit indirect, that either
the phase boundary has been crossed or that at 
least partial restoration of chiral symmetry with a concomitant
reduction of hadron masses has been achieved in the
hot and dense fireball created in the collision. 

\vspace{0.2cm}

Similar conclusions may be drawn from Figure 6, where we
have compared the above determined experimental freeze-out parameters
with the calculated position of the phase boundary. The surprising
result is that, even at freeze-out, the fireball parameters are very
close to those expected for a the quark-gluon plasma. Although the
system is, by definition, purely hadronic at freeze-out, its proximity
to the phase boundary suggests that the boundary was reached or even
crossed during the course of the evolution of the fireball towards
freeze-out. The arrows indicate where the systems should evolve from if
one takes predictions by cascade models such as RQMD as guideline for
the system parameters at the time of maximum compression and temperature.

\begin{figure}
\vspace*{-6.0cm}

\centerline{\psfig{figure=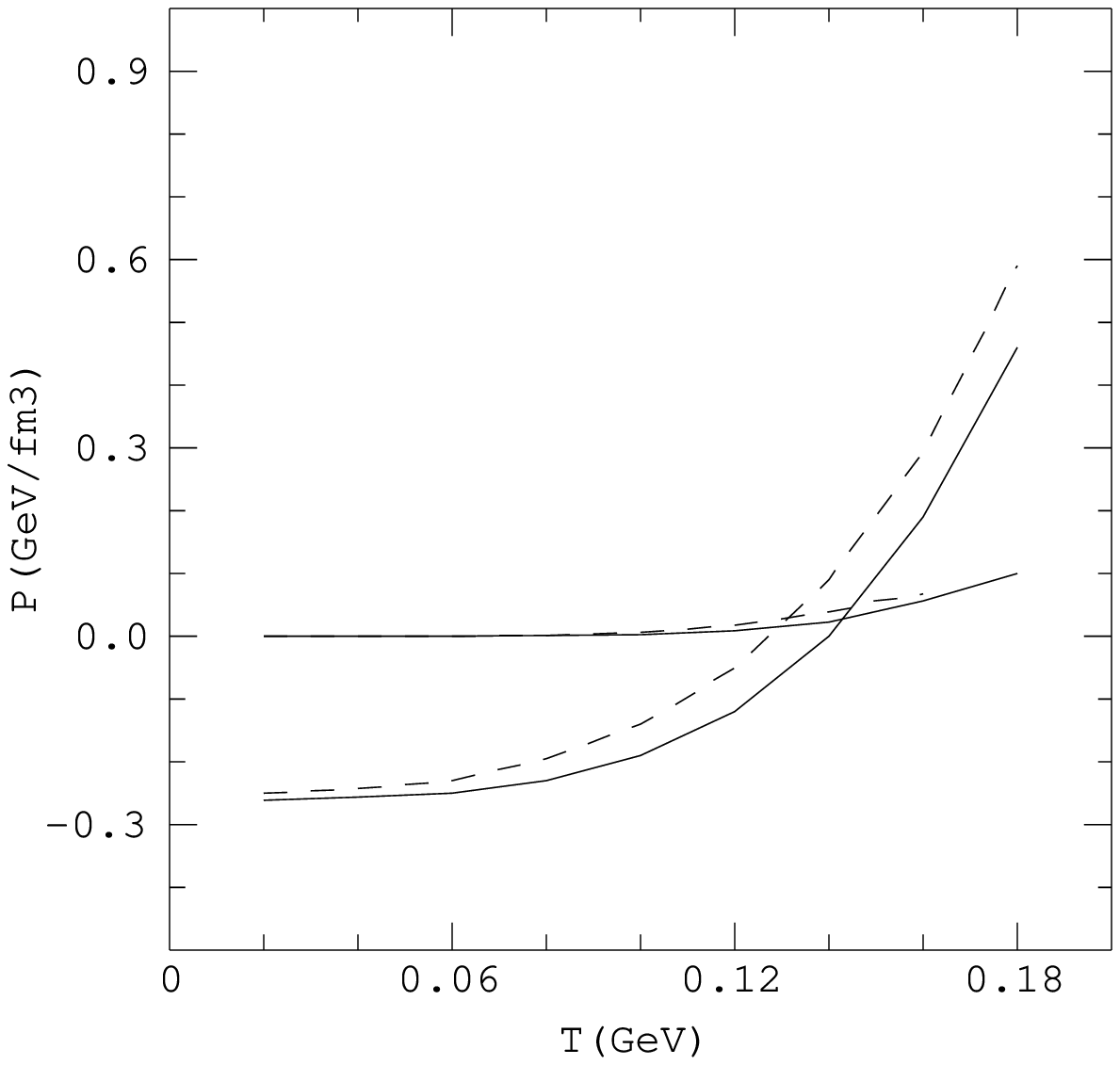,height=24cm}}

\caption{Dependence of pressure on temperature for the hadron gas
and quark-gluon plasma equations of state described in the text.
Solid (dashed) lines are for $\mu_B = 0.17 (0.54)$ GeV.
}
\label{fig1}
\end{figure}

\begin{figure}
\vspace*{-6.0cm}

\centerline{\psfig{figure=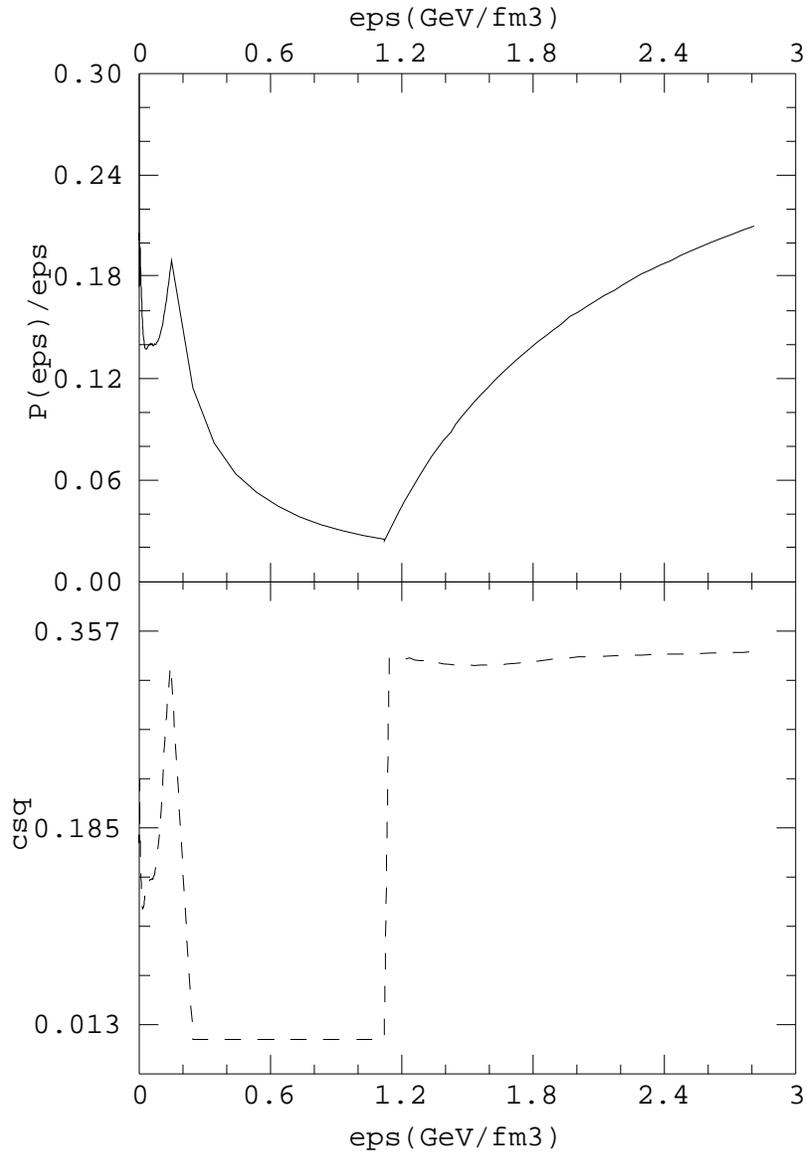,height=24cm}}


\caption{Top: $P/\epsilon$ {\it vs.} $\epsilon$
for $\mu_B = 0.54$ GeV. Bottom: the square of the speed of sound,
$c_s^2$ as function of the energy density $\epsilon$. Note that $c_s^2$
vanishes by construction in the mixed phase.
}
\label{fig2}
\end{figure}

\begin{figure}
\vspace*{-6.0cm}

\centerline{\psfig{figure=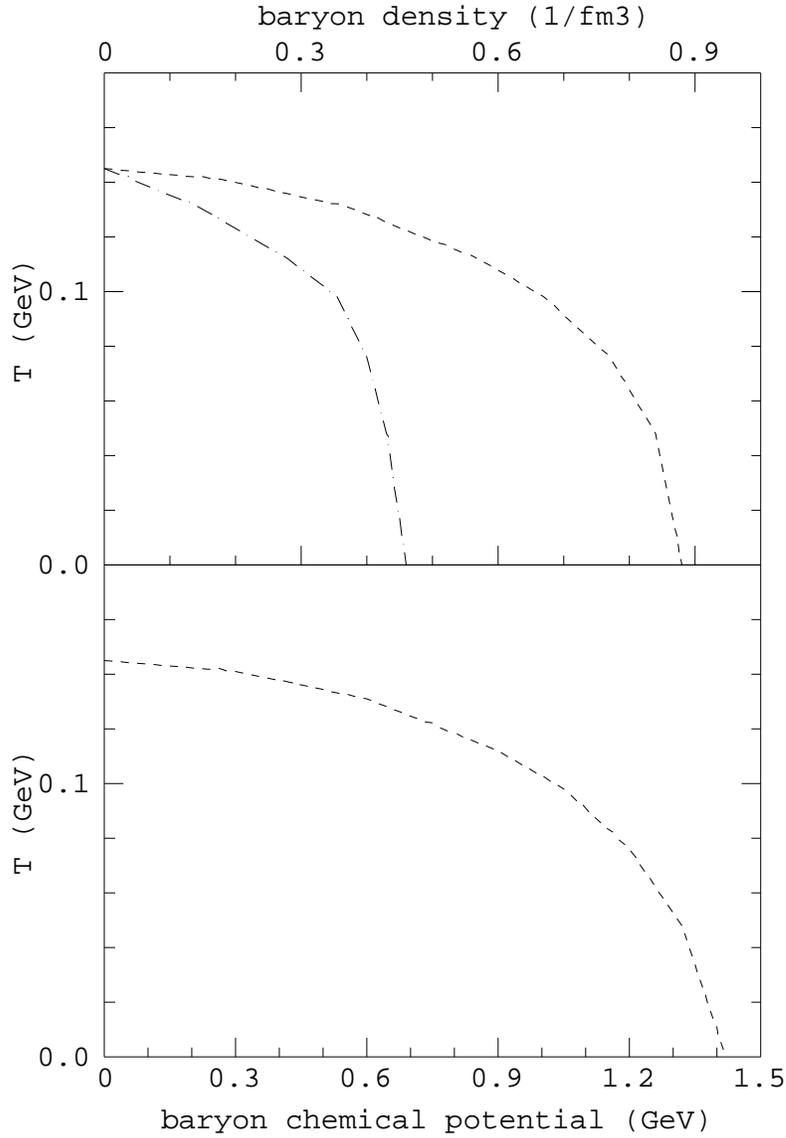,height=24cm}}


\caption{Phase boundary between hadronic and quark-gluon matter. Top: as
function of baryon density. 
Note the large jump in baryon density from hadronic (dash-dotted line)
to quark-gluon matter (dashed line). Bottom: as function of baryon
chemical potential. 
}
\label{fig3}
\end{figure}

\begin{figure}
\vspace*{-1.0cm}

\centerline{\psfig{figure=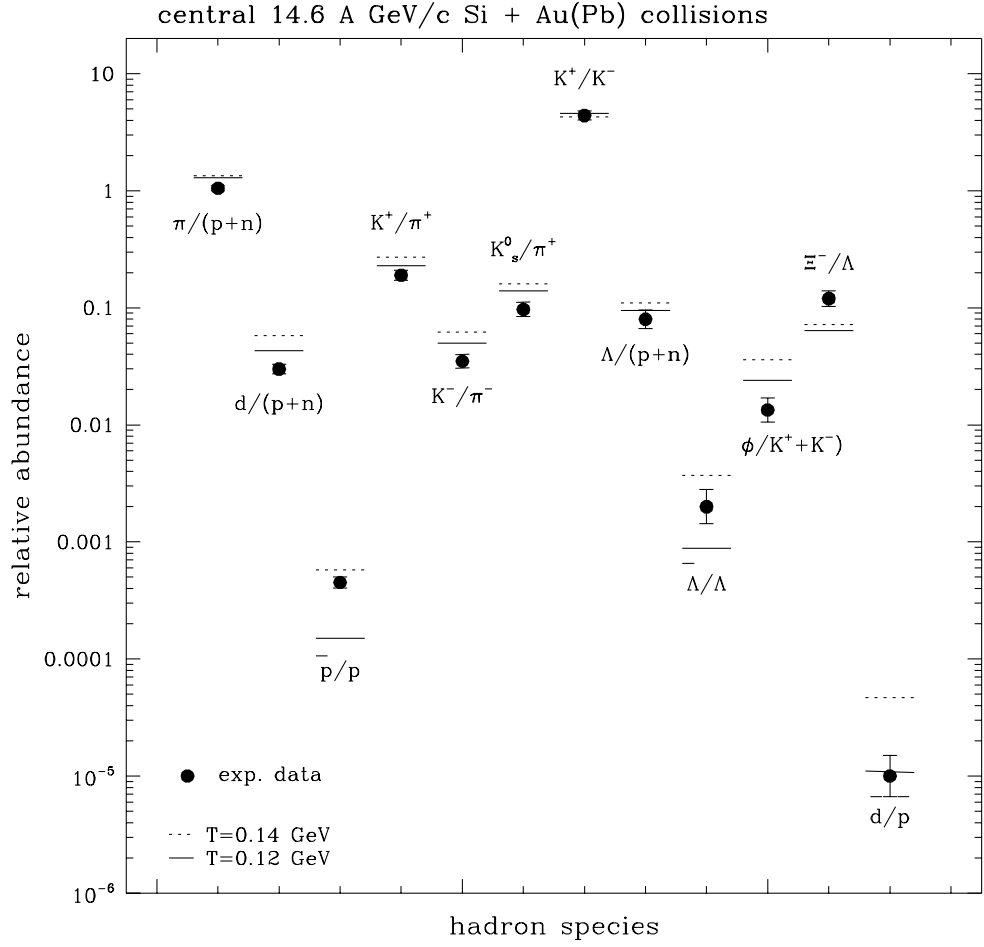,height=16cm}}

\caption{
Comparison of thermal model prediction to experimental data for hadron
abundance ratios at AGS energy. For details see text.
}
\label{fig4}
\end{figure}

\begin{figure}
\vspace*{-.6cm}

\centerline{\psfig{figure=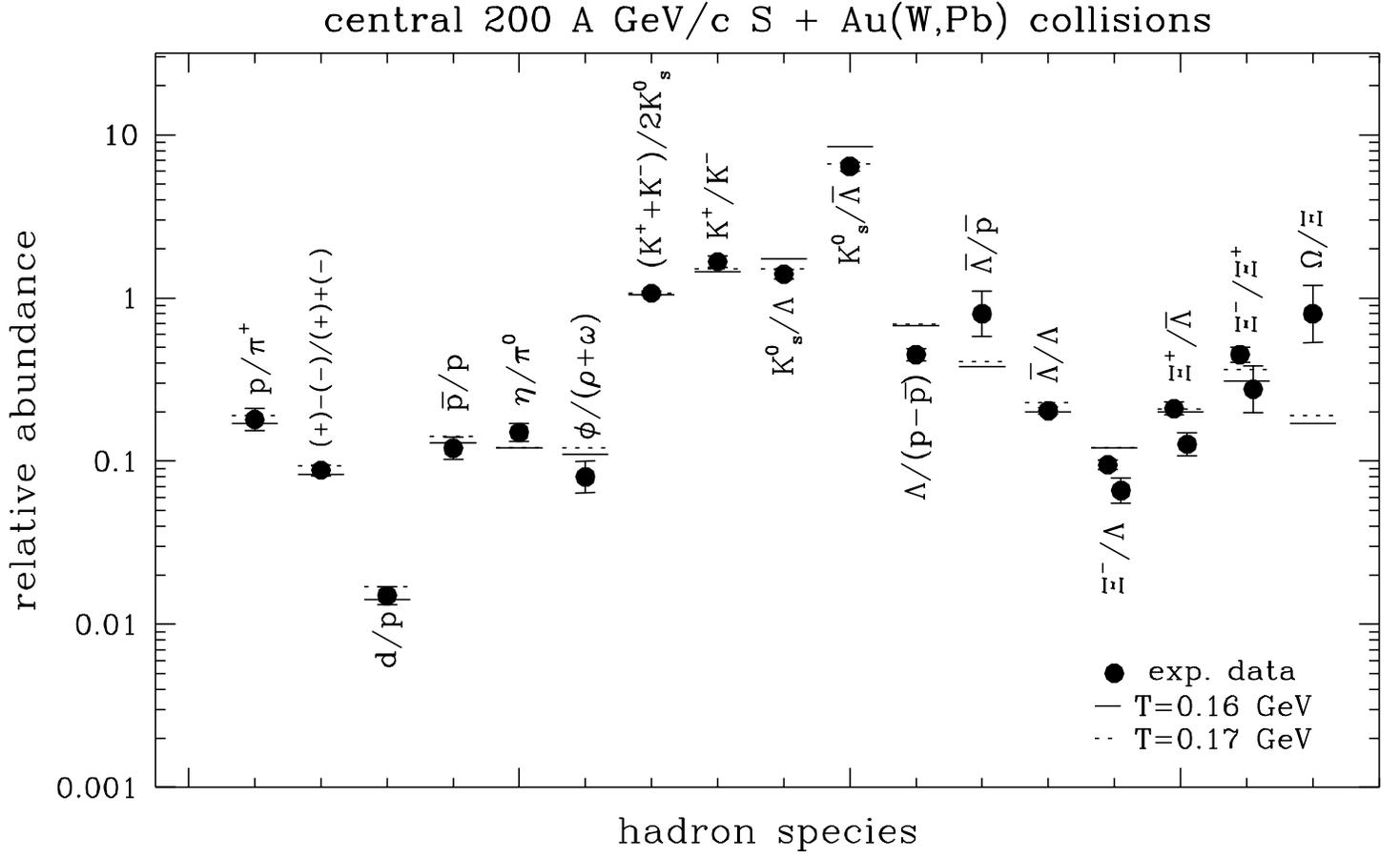,height=24cm}}

\vspace*{-6.0cm}

\caption{Comparison of thermal model prediction to experimental data for
hadron abundance ratios at SPS energy. The notation (+) or (-) refers to
the density of positively or negatively charged hadrons. For more
details see text. 
}
\label{fig5}
\end{figure}

\begin{figure}
\vspace*{-1.0cm}

\centerline{\psfig{figure=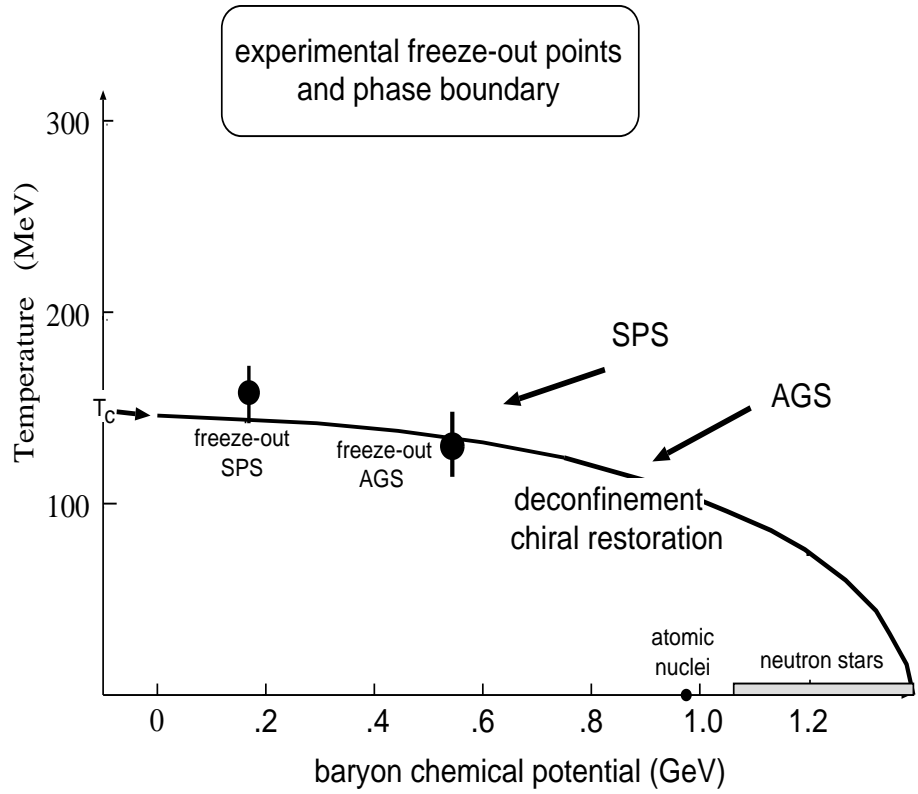,height=24cm}}

\vspace*{-9.0cm}

\caption{
Comparison of experimentally determined freeze-out parameters at AGS and
SPS energy with the phase boundary calculated in the present approach.}
\label{fig6}
\end{figure}

\centerline{\psfig{figure=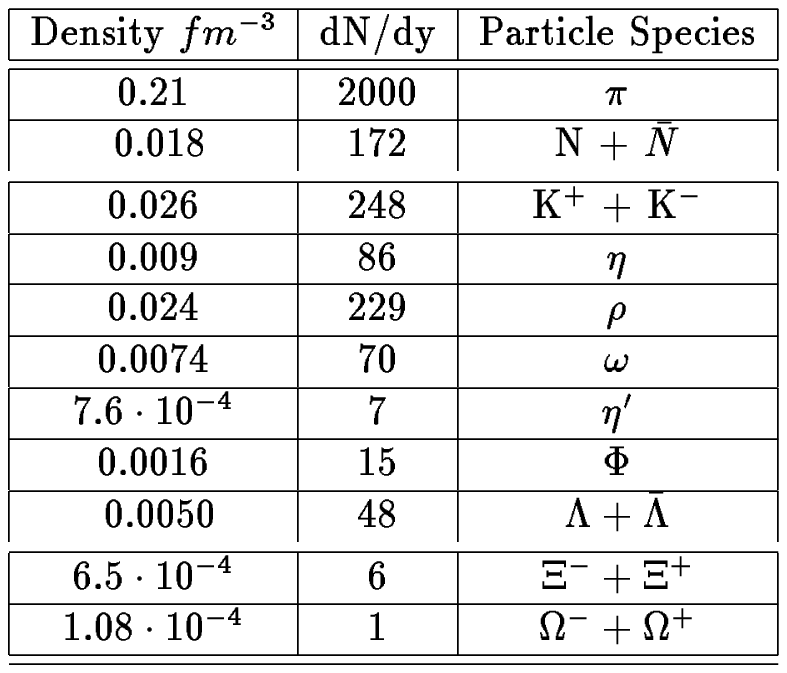,height=34cm}}

\vspace*{-20.0cm}

Table 1: Thermal model 
predictions for particle
and rapidity densities
at RHIC energy. For details see text.

\end{document}